\documentclass[12pt]{article}
\usepackage{geometry,amsmath,amssymb,graphicx,hyperref}
\newcommand{\Omicron}{\mathrm{O}}
\newcommand{\Zeta}{\mathrm{Z}}
\newcommand{\mathe}{\mathrm{e}}
\newcommand{\tmem}[1]{{\em #1\/}}
\newcommand{\tmop}[1]{\ensuremath{\operatorname{#1}}}
\newcommand{\tmtextbf}[1]{{\bfseries{#1}}}
\newcommand{\tmtextit}[1]{{\itshape{#1}}}
\newcommand{\tmtexttt}[1]{{\ttfamily{#1}}}
\begin{document}

\title{Simulating the All-Order Strong\\
Coupling Expansion V:\\
Ising Gauge Theory}
\author{
Tomasz Korzec\thanks{
e-mail: korzec@physik.hu-berlin.de} \; and
Ulli Wolff\thanks{
e-mail: uwolff@physik.hu-berlin.de}\\
Institut f\"ur Physik, Humboldt Universit\"at\\ 
Newtonstr. 15 \\ 
12489 Berlin, Germany
}
\date{}
\maketitle

\begin{abstract}
  We exactly rewrite the Z(2) lattice gauge theory with standard plaquette
  action as a random surface model equivalent to the untruncated set of its
  strong coupling graphs. We simulate such surfaces including Polyakov line defects
  that are moved by worm type update steps.
  Our Monte Carlo algorithms for the graph
  ensemble are reasonably efficient but not free of critical slowing down.
  Polyakov line correlators can be measured in this approach with small
  relative errors that are independent of the separation. As a first
  application our results are confronted with effective string theory
  predictions. In addition, the excess free energy due to twisted boundary
  conditions becomes an easily accessible observable. Our numerical
  experiments are in three dimensions, but the method is expected to work in
  any dimension.
\end{abstract}
\begin{flushright} HU-EP-12/52 \end{flushright}
\begin{flushright} SFB/CCP-12-98 \end{flushright}
\thispagestyle{empty}
\newpage

\section{Introduction}

This paper continues our series where we try to build on and generalize an
idea by Prokof'ev and Svistunov {\cite{prokofev2001wacci}} on alternate ways
of formulating and simulating simple statistical systems and Euclidean lattice
field theories. In our context, their idea has two aspects. One is that,
instead of summing over the field configurations in the original form, one
exactly rewrites the models as a finite or infinite sum over associated strong
coupling or high (classical) temperature graphs. The second aspect for spin
models is, that enlarging the class of such graphs from those of the partition
function to a larger set associated with the two point correlation function
makes simulations (`diagrammatic Monte Carlo') both simpler and much more
efficient with respect to critical slowing down. In {\cite{Wolff:2008km}} it
was shown that we may in addition exploit some freedom to design the enlarged
ensemble to our advantage and thus achieve an excellent signal to noise ratio
for interesting observables. It is this aspect that we will extend in this
paper to Polyakov line correlators in Abelian gauge theories while critical
slowing down is unfortunately not eliminated at the same time.

Some remarks are in order here. The standard usage of strong coupling
expansions is to evaluate a truncated series for some suitable observables and
then take the thermodynamic limit {\tmem{within this approximation}}. These
expansions in powers of some $\beta$ usually have finite radii of convergence
related to singularities or phase transitions. In systems with a large but
finite number of compact degrees of freedom however, these expansions
{\tmem{converge}} for many quantities like for instance the partition function
itself. Therefore, for such systems -- and we never simulate anything else --
the untruncated set of strong coupling graphs furnishes an {\tmem{exact}}
reformulation of the theory in which we can attempt to apply stochastic
summation methods. For a given action or Hamiltonian the graphs derived from
it reproduce finite lattice observables exactly and not just universal
features. For spin models we thus encounter representations in terms of loops
drawn on the lattice and the obvious generalization to gauge models studied
here consists of surfaces. The admissible graphs are restricted by
constraining rules, for instance on the number of lines that may touch at a
site, and we sum over such constrained objects. Solving the constraints in
terms of independent new variables would complete our reformulation to a
duality transformation, which is however not done in
{\cite{prokofev2001wacci}}. What the reformulations have in common is that
rather different observables may be advantageously computed in one or the
other. In models similar to Ising models duality {\cite{PhysRev.60.252}}
relates the strong coupling expansions in one model to the weak coupling
expansion in another. It is trivial, but amusing to remark, that a
diagrammatic Monte Carlo of the weak coupling (low temperature) expansion of
an Ising model is nothing but a standard simulation, as spin-flip related
pairs of configurations correspond to one naive weak coupling graph. In the
systematic weak coupling approximation one just truncates these contributions
guided by the size of their Boltzmann weights.

Our previous work in this series extended the method from the Ising model
{\cite{Wolff:2008km}} to the O($N$), CP($N - 1$) models and to fermions, see
{\cite{Wolff:2010zu}} for an overview and more references. For fermions an
unsolved sign problem hampers simulations in more than two dimensions while in
the other cases the technique is expected to be efficient in arbitrary
dimension. Studies of O($N$) as a special case of certain loop graph models on
particular lattices are also found in {\cite{Liu:2010uw}},
{\cite{Liu:2012ca}}.

There have been other recent attempts to apply the `worm' method to Abelian
gauge models {\cite{Endres:2006xu}}, {\cite{Azcoiti:2009md}}, {\cite{Korzec:2010sh}},
{\cite{Gattringer:2012jt}}, 
{\cite{Delgado:2012tm}}. The important new feature in the present study is in
our mind however, that we succeed in generalizing the low noise estimators for
fundamental correlations from spin to gauge systems. Also the r\^ole of
twisted boundary conditions on the torus will be discussed which leads to
interesting topological universal finite size observables.

As we get access to precise results for Polyakov line correlators, a
comparison with low energy effective string models suggests itself as a first
interesting application of our simulation technique. Such studies form an
active research area and we here offer a demonstration of the usefulness of
our method in the context. A very profound such study is beyond the scope of
this publication. The field has a long history with a boost, as far as we see,
after the papers by L\"uscher and Weisz {\cite{Luscher:2002qv}},
{\cite{Luscher:2004ib}} which interestingly were also triggered by algorithmic
improvement. The low energy description of large Wilson and Polyakov loops in
confining gauge theories was pushed to the two loop level in the spirit of a
general effective field theory approach reminiscent of chiral perturbation
theory vis a vis QCD. It was found that to this order and in three dimensions
and taking into account all symmetries, no free low energy coupling constants
enter. This investigation was later even extended to the three loop level in
{\cite{Aharony:2009gg}} so that we have remarkable absolute predictions to
check with our low noise long distance results for the Polyakov loop
correlator.

For the Z(2) Ising gauge theory in three dimensions that we adopt here as a
test case, there actually exists an enormous body of precise results in the
literature, see {\cite{Caselle:2012rp}}, {\cite{Billo:2011fd}} for recent
papers with further references. Here the duality of the gauge theory with the
spin model is exploited and the simulations are conducted there. The method is
described in detail in {\cite{Caselle:2002ah}}. A Wilson loop in the gauge
theory translates into a ratio of partition functions in the spin model
differing by flipped bonds. \ Such a ratio is factorized into many ratios
differing in single bonds, which are evaluated in separate simulations. While
this is an intrinsically three dimensional method, the same is not true for
our more direct approach, although we here test it in $D = 3$ as well.

The paper is organized as follows. In Section 2. we introduce the model, its
boundary conditions, rewrite it as a surface ensemble and collect some
formulae of the effective string description. This is followed by the
development of our Monte Carlo method for the surface ensemble in Section 3.
and a report of our numerical experiments in Section 4. We end on conclusions
in Section 5. and an Appendix listing numerical data.

\section{Z(2) gauge theory}

In this section we setup our model on a hypercubic lattice in $D$ Euclidean
dimensions. We include the definition of fluctuating twisted boundary
conditions. Then we introduce the one-to-one reformulation as a surface model,
which generalizes the loop (re)formulation of spin models.

\subsection{Twisted boundary conditions}

We consider a gauge field $\sigma (x, \mu) \equiv \sigma_{\mu} (x) = \pm 1 \in
\Zeta (2)$ defined on the links of a $D$-dimensional hypercubic periodic
lattice of extent $L_{\mu}$ in directions $\mu = 0, 1, \ldots, D - 1$. The
standard Wilson action is defined on plaquettes $(x, \mu < \nu)$ by
\begin{equation}
  - S [\sigma, P] = \beta \sum_{x, \mu < \nu} P \left( x, \mu, \nu \right)
  \sigma (x, \mu) \sigma (x + \hat{\mu}, \nu) \sigma (x + \hat{\nu}, \mu)
  \sigma (x, \nu) .
\end{equation}
The plaquette dependent background field $P \left( x, \mu, \nu \right) \in
\Zeta (2)$ will be useful later and may be first imagined to be unity until
further notice. Generalized periodic boundary conditions{\footnote{A fruitful
point of view here is to consider all fields on the {\tmem{infinite}} lattice
with a {\tmem{finite}} subset of independent variables as all others are
`locked' by periodicity.}} demand that the gauge field is periodic up to gauge
transformations{\footnote{Gauge invariant densities will thus be periodic in
the usual sense. This generalizes antiperiodic boundary conditions in the
Ising model which leave Z(2) symmetric composites periodic.}}
{\cite{'tHooft:1979uj}}
\begin{equation}
  \sigma (x + L_{\nu} \hat{\nu}, \mu) = \sigma (x, \mu) \tau_{\nu} \left( x
  \right) \tau_{\nu} \left( x + \hat{\mu} \right)
\end{equation}
specified by fixed `external' transition functions $\tau_{\nu} \left( x
\right)$. Because the shifts form an Abelian group it is necessary that shifts
$x \rightarrow x + L_{\nu} \rightarrow x + L_{\nu} + L_{\lambda}$ come with
the same gauge transformation as the double shift in the opposite order, which
however still leaves the possibility
\begin{equation}
  \tau_{\nu} \left( x \right) \tau_{\lambda} ( x + L_{\nu} \hat{\nu}
  ) = \gamma_{\nu \lambda} \tau_{\lambda} \left( x \right) \tau_{\nu}
  ( x + L_{\lambda} \hat{\lambda} ), \hspace{1em} \gamma_{\nu
  \lambda} = \gamma_{\lambda \nu} \in \Zeta \left( 2 \right) .
\end{equation}
The case $\gamma_{\nu \lambda} = - 1$ is possible because the action of
$\tau_{\nu}$ on gauge fields is independent of the global sign of the
transition function. Then we have twisted boundary conditions in the $\nu
\lambda$ plane while we call planes with $\gamma_{\nu \lambda} = + 1$
untwisted or just periodic below.

If we gauge transform
\begin{equation}
  \sigma \left( x, \mu \right) \rightarrow \sigma \left( x, \mu \right) \rho
  \left( x \right) \rho \left( x + \hat{\mu} \right) \label{gtrho}
\end{equation}
in the ordinary sense, this implies the following change of the transition
functions $\tau_{\nu}$
\begin{equation}
  \tau_{\nu} \left( x \right) \rightarrow \tau_{\nu} \left( x \right) \rho
  \left( x \right) \rho \left( x + L_{\nu} \hat{\nu} \right),
\end{equation}
they transform like parallel transporters on a set of `super-lattices' with
spacings $L_{\mu}$. The $\gamma_{\mu \nu}$ represent the gauge invariant
content of the transition functions. \ For a given set of twists $\gamma_{\mu
\nu}$ we now define reference transition functions
\begin{equation}
  \tau^{\left( \gamma \right)}_{\nu} \left( x \right) = \prod_{\lambda < \nu}
  \left( \gamma_{\nu \lambda} \right)^{\lceil x_{\lambda} / L_{\lambda}
  \rceil}
\end{equation}
where we round upwards to an integer (`ceil') in the exponent. Then the
product $\tau_{\nu} \left( x \right) \tau^{\left( \gamma \right)}_{\nu} \left(
x \right)$ has trivial twist and can be gauged to unity. Hence we may assume
that the transition functions have the form $\tau^{\left( \gamma \right)}$ and
then the gauge field is periodic except (possibly) for
\begin{equation}
  \sigma (x + L_{\nu} \hat{\nu}, \mu) = \gamma_{\mu \nu} \sigma (x, \mu)
  \hspace{1em} \tmop{if} \hspace{1em} \mu < \nu \hspace{1em} \tmop{and}
  \hspace{1em} x_{\mu} = 0 \hspace{1em} \left( \tmop{mod} L_{\mu} \right) .
  \label{persig}
\end{equation}
By a further change of variables, we may arrive at fully periodic $\sigma (x,
\mu)$ again if we absorb signs into
\begin{equation}
  P_{\gamma} \left( x, \mu, \nu \right) = \left( \gamma_{\mu \nu}
  \right)^{\delta_{x_{\mu}, 0} \delta_{x_{\nu}, 0}}
\end{equation}
with {\tmem{periodic}} $\delta$ symbols (period $L_{\mu}, L_{\nu}$
respectively). Thus for each twisted plane we have a $D - 2$ dimensional set
('stack') of negative plaquettes that represent a background flux or disorder
(`vortices') concentrated on a line, sheet, ... for $D = 3, 4, \ldots$. This
step is analogous to absorbing antiperiodic boundary conditions into a
background gauge field, see {\cite{Hogervorst:2011}} for example.

\subsection{Gauge theory as a surface model}

Next we introduce the partition function with current insertions
\begin{equation}
  \tilde{Z}_{\gamma} [j] = 2^{- N_l} \sum_{\sigma} \mathe^{- S [\sigma,
  P_{\gamma}]}  \prod_{x, \mu} \sigma (x, \mu)^{j (x, \mu)} . \label{Ztj}
\end{equation}
In this formula we sum over strictly periodic $\sigma (x, \mu) = \pm 1$
independently at all $0 \leqslant x_{\mu} < L_{\mu}$ and $\mu$, and $j (x,
\mu) \in \{0, 1\}$ is a periodic external field. The number of independent
sites, links, plaquettes and 3-cubes (for later use) is
\begin{equation}
  N_x = \prod_{\mu} L_{\mu}, \hspace{1em} N_l = D N_x, \hspace{1em} N_p = N_l
  \left( D - 1 \right) / 2, \hspace{1em} N_c = N_p \left( D - 2 \right) / 3.
\end{equation}
Performing a local gauge change of variables (\ref{gtrho}) with periodic $\rho
\left( x \right)$ one derives
\begin{equation}
  \tilde{Z}_{\gamma} [j] = \tilde{Z}_{\gamma} [j] \prod_x \rho
  (x)^{\partial^{\ast}_{\mu} j_{\mu} (x)}
\end{equation}
with the divergence
\begin{equation}
  \partial^{\ast}_{\mu} j_{\mu} (x) \equiv \sum_{\mu} \left[ j_{\mu} (x) +
  j_{\mu} (x - \hat{\mu}) \right],
\end{equation}
being the sum over the $2 D$ links surrounding $x$. As $\rho$ can be arbitrary
this shows that only divergence free currents (in the Z(2) sense), for which
\begin{equation}
  \partial^{\ast}_{\mu} j_{\mu} (x) = 0 \hspace{1em} \left( \tmop{mod} 2
  \right) \label{locflux}
\end{equation}
holds at all sites, yield nonzero $\tilde{Z}_{\gamma} [j]$. In addition we may
flip the gauge field $\sigma_{\mu}$ on any $D - 1$ dimensional hyperplane
$x_{\mu} = z$ orthogonal to the $\mu$-direction (`layers'). A similar argument
as before yields the requirement that the layer sums
\begin{equation}
  \sum_{x, x_{\mu} = z} j (x, \mu) = 0 \hspace{1em} \left( \tmop{mod} 2
  \right) \hspace{1em} 0 \leqslant z < L_{\mu} \label{globflux}
\end{equation}
must be even for all layers, which are hence pierced by an even number of
current quanta. Equivalently we may say that the link field $j \left( x, \mu
\right)$ must have vanishing Z(2) winding number with respect to all torus
directions.

Using
\begin{equation}
  \mathe^{\beta \sigma} = \cosh (\beta) \sum_{n = 0, 1} t^n \sigma^n,
  \hspace{1em} t = \tanh (\beta) \label{locWil}
\end{equation}
for each plaquette, we introduce a field $n (x, \mu, \nu) \in \{0, 1\}$. Then
we may average over the original $\sigma (x, \mu)$ which leave behind
constraints only. We arrive at
\begin{equation}
  Z_{\gamma} [j] = \sum_n t^{\sum_{x, \mu < \nu} n (x, \mu, \nu)}
  \Phi_{\gamma} \left[ n \right] \delta [\partial_{\mu}^{\ast} n_{\mu \nu} +
  j_{\nu}], \hspace{1em} \tilde{Z}_{\gamma} [j] = (\cosh \beta)^{N_p}
  Z_{\gamma} [j] . \label{Zj}
\end{equation}
The sign is given by
\begin{equation}
  \Phi_{\gamma} \left[ n \right] = \prod_{x, \mu < \nu} \left[ P_{\gamma}
  \left( x, \mu, \nu \right) \right]^{n \left( x, \mu, \nu \right)} =
  \prod_{\mu < \nu} \left( \gamma_{\mu \nu} \right)^{w_{\mu \nu} \left[ n
  \right]},
\end{equation}
where for each $n \left( x, \mu, \nu \right)$ configuration we have introduced
the `wrapping' numbers
\begin{equation}
  w_{\mu \nu} = \sum_{x \left|_{x_{\mu} = x_{\nu} = 0} \right.} n \left( x,
  \mu, \nu \right) \hspace{1em} \left( \tmop{mod} 2 \right), \hspace{1em}
  w_{\mu \nu} \in \left\{ 0, 1 \right\} . \label{wdef}
\end{equation}
They are topological quantities{\footnote{We define it here also for
configurations with $j \neq 0$. A topological meaning, independent of our
choice of $\tau^{\left( \gamma \right)}$, is given however for vacuum
configurations only.}} that count (modulo two) how many times the surfaces
made of plaquettes with $n \left( x, \mu, \nu \right) = 1$ wind around planes
and represent two dimensional generalizations of the winding numbers of Ising
loops. To define the divergence of the plaquette field, we
extend{\footnote{Note that for mod 2 additive variables `symmetric' and
`antisymmetric' coincides.}}
\begin{equation}
  n (x, \mu, \nu) = n (x, \nu, \mu), \hspace{1em} n (x, \mu, \mu) = 0.
\end{equation}
The constraint (mod 2) combines the $2 (D - 1)$ plaquettes surrounding each
link
\begin{equation}
  \delta [\partial_{\mu}^{\ast} n_{\mu \nu} + j_{\nu}] = \prod_{x, \nu}
  \delta_{\partial_{\mu}^{\ast} n_{\mu \nu} (x), j_{\nu} (x)} .
\end{equation}
We note that consistently this constraint can only be satisfied by $j$ that
obey (\ref{locflux}) and (\ref{globflux}). The proof looks slightly funny, for
example
\[ \partial_{\nu}^{\ast} j_{\nu} = \partial_{\mu}^{\ast} \partial_{\nu}^{\ast}
   n_{\mu \nu} = 2 \sum_{\mu < \nu} \partial_{\mu}^{\ast}
   \partial_{\nu}^{\ast} n_{\mu \nu} \hspace{1em} \Rightarrow \hspace{1em}
   \partial_{\nu}^{\ast} j_{\nu} = 0 (\tmop{mod} 2) . \]
An easy to interpret case is a configuration $j \equiv 0$. Then the plaquettes
$n (x, \mu, \nu) = 1$ form a surface which may branch and consist of
disconnected components. The zero divergence condition demands that they are
closed and have no boundaries, each link is surrounded by an even number of
surface elements. Boundaries arise if $j$ is nonzero and, due to
(\ref{locflux}) and (\ref{globflux}) they may be written as a superposition of
contractable closed current loops. It will be helpful for the reader to work
out the closely analogous but geometrically simpler loop formulation of the
Ising spin model in this language.

In analogy to the steps taken for spin models we now form the current ensemble
\begin{equation}
  \mathcal{Z}= \sum_j R^{- 1} [j] Z [j] = \sum_n t^{\sum_{x, \mu < \nu} n (x,
  \mu, \nu)} R^{- 1} [\partial_{\mu}^{\ast} n_{\mu \nu}] \label{ZZj}
\end{equation}
where the non-negative weight $R^{- 1} [j]$ will be specified later and $Z$
without subscript stands for trivial twist $Z = Z_{\gamma \equiv 1}$. Note
that we here include all wrapping numbers in the sum over $n$ with no extra
signs, which corresponds to trivial twist in all planes.

Expectation values in this ensemble are given by
\begin{equation}
  \langle \langle \mathcal{O}[n] \rangle \rangle = \frac{1}{\mathcal{Z}}
  \sum_n \mathcal{O} [n] t^{\sum_{x, \mu, \nu} n (x, \mu, \nu)} R^{- 1}
  [\partial_{\mu}^{\ast} n_{\mu \nu}]. \label{mean}
\end{equation}
In addition we define `vacuum' expectation values
\begin{equation}
  \langle \langle \mathcal{O}[n] \rangle \rangle_0 = \frac{\langle \langle
  \mathcal{O}[n] \delta [\partial_{\mu}^{\ast} n_{\mu \nu}] \rangle
  \rangle}{\langle \langle \delta [\partial_{\mu}^{\ast} n_{\mu \nu}] \rangle
  \rangle} \label{mean0}
\end{equation}
on the subset of defect-free configurations. Such quantities do not depend on
the choice of $R$. It is now obvious that ratios of partition functions of
different twist, which lead to interesting observables, are given by
topological observables
\begin{equation}
  \frac{Z_{\gamma}}{Z} = \langle \langle \prod_{\mu < \nu} \left( \gamma_{\mu
  \nu} \right)^{w_{\mu \nu} \left[ n \right]} \rangle \rangle_0 .
  \label{wgamma}
\end{equation}

We shall later see that there are Monte Carlo algorithms that are ergodic only
in the sector of trivial wrapping in some or even all planes. It is clear now
that the corresponding ensembles can be considered as arising by dynamically
averaging over twisted and untwisted boundary conditions for these planes
(fluctuating boundary conditions).

By differentiating $\mathcal{Z} \langle \langle \delta [\partial_{\mu}^{\ast}
n_{\mu \nu}] \rangle \rangle$ with respect to $\beta$ we obtain the relation
between the average plaquette $E$ of the original theory and the total surface
area in vacuum configurations
\begin{equation}
  E = t + (t^{- 1} - t) \frac{1}{N_p} \sum_{x, \mu < \nu} \left\langle
  \left\langle n (x, \mu, \nu) \right\rangle \right\rangle_0 . \label{E0}
\end{equation}
Below we shall find that close to the critical point in $D = 3$ we have the
rather high values $\langle \langle n \rangle \rangle_0 \approx 1 / 3$ and $E
\approx 0.95$.

\subsection{Polyakov loop ensemble}

The source $j \left( x, \mu \right)$ allows to place a large variety of defect
configurations like arbitrary Wilson loops. We here define however a highly
restricted framework involving only two Polyakov lines in the 0-direction that
are located at $\vec{u} = \left( u_1, \ldots, u_{D - 1} \right)$ and an
analogous $\vec{v}$. The corresponding conserved current is
\begin{equation}
  j^{\left( \vec{u}, \vec{v} \right)} \left( x, \mu \right) = \delta_{\mu, 0}
  \left[ \delta_{\vec{u}, \vec{x}} + \delta_{\vec{v}, \vec{x}} \right]
  \hspace{1em} \left( \tmop{mod} 2 \right) . \label{juv}
\end{equation}
For coinciding $\vec{u} = \vec{v}$ it vanishes and there is no defect.

We consider the ensemble
\begin{eqnarray}
  \mathcal{Z} & = & \sum_{\vec{u}, \vec{v}} \rho^{- 1} \left( \vec{u} -
  \vec{v} \right) Z [j^{\left( \vec{u}, \vec{v} \right)}] \nonumber\\
  & = & \sum_{n, \vec{u}, \vec{v}} \rho^{- 1} \left( \vec{u} - \vec{v}
  \right) t^{\sum_{x, \mu, \nu} n (x, \mu, \nu)} \delta [\partial_{\mu}^{\ast}
  n_{\mu \nu} - j^{\left( \vec{u}, \vec{v} \right)}_{\nu}] .  \label{Polens}
\end{eqnarray}
Here the weight $R [j^{\left( \vec{u}, \vec{v} \right)}]$ has been specialized
to $0 < \rho \left( \vec{u} - \vec{v} \right) = \rho \left( \vec{v} - \vec{u}
\right)$ while $R^{- 1} \left[ j \right]$ vanishes if $j$ is not of the form
(\ref{juv}).

The two point function (in the original gauge theory) of the Polyakov loop
operator
\begin{equation}
  \pi \left( \vec{x} \right) = \prod_{x_0 = 0}^{L_0 - 1} \sigma_0 \left( x
  \right) \label{Polop}
\end{equation}
is now given by a ratio of expectation values of counters
\begin{equation}
  G \left( \vec{x} \right) = \langle \pi \left( \vec{x} \right) \pi (
  \vec{0} ) \rangle = \rho \left( \vec{x} \right) \frac{\langle
  \langle \delta_{\vec{x}, \vec{u} - \vec{v}} \rangle \rangle}{\langle \langle
  \delta_{\vec{u}, \vec{v}} \rangle \rangle} \label{Gpol}
\end{equation}
if we normalize $\rho(\vec{0}) = 1$.

\subsection{Contact with transfer matrices, effective string theory}

The Polyakov line correlator is the ratio of partition functions with and
without two static charges and thus given by
\begin{equation}
  G \left( \vec{x} \right) = w_n \sum_{n \geqslant 0} \mathe^{- V_n \left(
  \vec{x} \right) L_0},
\end{equation}
where $\mathe^{- V_n \left( \vec{x} \right)}$ labels the corresponding
eigenvalues of the transfer matrix in the 0-direction with charges separated
by $\vec{x}$ and integer weights $w_n$ account for degeneracies
{\cite{Luscher:2004ib}}.

Alternatively we may consider a transfer matrix in a spatial direction, $k =
1$ for example. Then the Polyakov line operator excites a flux state whose
energy strongly depends on its length $L_0$ (now a `transverse' direction) and
we may consider the correlation
\begin{equation}
  C (y) = \sum_{\vec{x} \left|_{x_1 = y} \right.} G \left( \vec{x} \right) =
  \sum_{n \geqslant 0} |v_n |^2 \mathe^{- \tilde{E}_n y} \hspace{1em} (y
  \ll L_1), \label{Ccor}
\end{equation}
where we have projected to zero momentum for the $k > 1$ directions and $v_n$
is a nontrivial matrix element in this case.

In effective string theories one can compute the energies $\tilde{E}_n$ as an
asymptotic expansion in $1 / L_0$. On the basis of the Nambu-Goto action for
example one predicts for $D = 3$ the relation {\cite{Arvis:1983fp}}
{\cite{Luscher:2004ib}}
\begin{equation}
  z^2 = s^2 \left( 1 - \frac{1}{3 s} \right), \hspace{1em} s = \frac{\sigma
  L_0^2}{\pi}, \hspace{1em} z = \frac{\tilde{E}_0 L_0}{\pi} \label{NGrel}
\end{equation}
for the ground state energy which is expected to be relevant for $z, s
\rightarrow \infty$, and $\sigma$ is the (zero temperature) string tension.
Formula (\ref{NGrel}) implies asymptotic expansions
\begin{equation}
  \tilde{E}_0 = \sigma L_0 - \frac{\pi}{6 L_0} - \frac{\pi^2}{72 \sigma L_0^3}
  - \frac{\pi^3}{432 \sigma^2 L_0^5} + \Omicron \left( L_0^{- 7} \right)
  \label{E0ex}
\end{equation}
and, expanding $s$ in $z^{- 1}$,
\begin{equation}
  \sigma = \frac{\tilde{E}_0}{L_0} + \frac{\pi}{6 L_0^2} + \frac{\pi^2}{72
  \tilde{E}_0 L_0^3} + \frac{0}{\tilde{E}_0^2 L_0^4} + \Omicron \left(
  \tilde{E}_0^{- 3} L_0^{- 5} \right) . \label{sigex}
\end{equation}
It is interesting to note that in general the Nambu Goto picture is not
expected to hold to all orders in the asymptotic expansion, but the terms
exhibited above have been uniquely derived for {\tmem{general}} effective low
energy actions just restricted by symmetries {\cite{Luscher:2004ib}},
{\cite{Aharony:2009gg}}.

In addition it is shown \ in {\cite{Luscher:2004ib}} that, up to rotational
symmetry breaking cutoff effects, the original correlation is given in terms
of the $\tilde{E}_n$ by the expansion
\begin{equation}
  G \left( \vec{x} \right) = \sum_{n \geqslant 0} |v_n |^2 2 r \left(
  \frac{\tilde{E}_n}{2 \pi r} \right)^{\frac{1}{2} \left( D - 1 \right)}
  K_{\frac{1}{2} \left( D - 3 \right)} \left( \tilde{E}_n r \right)
  \label{GLW} 
\end{equation}
with $r = \left| \vec{x} \left| \right. \right.$ and the modified Bessel
function $K_{\nu}$. Note that here an infinite volume is assumed.

The estimator for $C \left( y \right)$ in our extended ensemble (\ref{Polens})
follows from (\ref{Gpol}) and is given by the simple observable
\begin{equation}
  C \left( y \right) = \frac{\langle \left\langle \delta_{y, u_1 - v_1} \rho
  \left( \vec{u} - \vec{v} \right) \right\rangle \rangle}{\langle \langle
  \delta_{\vec{u}, \vec{v}} \rangle \rangle} \label{Cz}
\end{equation}
that simply follows from the statistics of the defect-line separations. Note
that this result is independent of the choice of $\rho$ which we shall hence
optimize with regard to the numerical simulation of (\ref{Polens}). In
practice one will symmetrize over spatial directions if they are symmetric
with respect to extent and boundary conditions.

\section{Simulation algorithms}

We now introduce several algorithms to simulate the surface representation of
Z(2) lattice gauge theory.

\subsection{Defect conserving update}

A very simple update step consists of the following sequence CF (`cube flip')
of operations{\footnote{Such steps for U(1) have been introduced in
{\cite{Sterling:1983fs}}.}}
\begin{itemize}
  \item take a 3-cube $c \equiv x, \mu < \nu < \lambda$,
  
  \item propose to flip all 6 plaquettes associated with this cube which is
  the set
  \[ \mathcal{P} \left( x, \mu, \nu, \lambda \right) = \]
  \[ \left\{ ( x, \mu, \nu ), ( x + \hat{\lambda}, \mu, \nu
     ), \left( x, \mu, \lambda \right), \left( x + \hat{\nu}, \mu,
     \lambda \right), \left( x, \nu, \lambda \right), \left( x + \hat{\mu},
     \nu, \lambda \right) \right\} \]
  i.e. $n \left( p \right) \rightarrow 1 - n \left( p \right)$ for all $p \in
  \mathcal{P}$,
  
  \item accept this proposal with the Metropolis probability $\min \left( 1, q
  \right)$ with
  \[ q = \prod_{p \in \mathcal{P}} t^{1 - 2 n \left( p \right)} . \]
\end{itemize}
Such steps may be iterated either with randomly chosen $c$ or as a systematic
sweep covering each $c$ once in some order. This can be shown to lead to an
ergodic algorithm at fixed $j$ (vacuum graphs with $j \equiv 0$ for example)
and within a {\tmem{fixed wrapping number}} sector. We expect (and confirm)
however dynamical exponents close to two for such local updates that preserve
the vacuum graph constraints. This is similar to just flipping links around
plaquettes in the strong coupling loop representation of Ising spin models.

A natural generalization of the worm algorithm {\cite{prokofev2001wacci}}
would be to allow to `open up' vacuum graphs and allow excursions to the
enlarged state space of (allowed) $j \neq 0$ (possibly traversing `useful'
configurations regarding observables of interest) and eventually returning to
a vacuum graph. We have a long record of not really successful experiments of
this type, see {\cite{Korzec:2010sh}} for experiments with U(1) gauge theory.
In Z(2) we have simulated correct ergodic algorithms where we included a chemical
potential per defect link $j \left( x, \mu \right) = 1$ to control their
number. We could find values that have led to ensembles of randomly shaped
(typically irregular) Wilson loops where also vacuum configurations
(re-)appeared at a reasonable rate. None of these attempts has so-far led
however either to fast dynamics or to easily interpretable observables.
Therefore we come up with the hybrid below which combines cube flips with
Polyakov line pair defects only where we have to tolerate however some
critical slowing down.

\subsection{Polyakov line moving update}\label{PSsubsec}

We now describe worm-type updates referring to the ensemble (\ref{Polens}). If
(say) $\vec{u}$ is moved to a neighboring site $\vec{u}'$ in the spatial
direction $i$ the corresponding line defect moves and the whole `ladder' of
$L_0$ plaquettes in the $0 i$ plane `above' this spatial link is flipped if
the move is accepted. Such a move clearly preserves the constraint. To specify
the procedure in detail it is helpful to define the auxiliary spatial link
field
\begin{equation}
  k \left( \vec{x}, i \right) = \sum_{z = 0}^{L_0 - 1} n \left( \vec{x} + z
  \hat{0}, 0, i \right) \in \left[ 0, L_0 \right) .
\end{equation}
Now we execute the following sequence PS ('Polyakov shift') of (standard worm)
steps:
\begin{itemize}
  \item If $\vec{u} = \vec{v}$ holds, we randomly re-locate both together to a
  new site on the $D - 1$ dimensional lattice.
  
  \item Pick one of the $2 \left( D - 1 \right)$ spatial neighbors $\vec{u}'$
  of $\vec{u}$, such that $\vec{u}' = \vec{u} \pm \hat{i}$.
  
  \item Accept the proposal with the probability (pre-tabulated, of course)
  \begin{equation}
    \tilde{p}_{i, +} = \min \left[ 1, t^{\left[ L_0 - 2 k \left( \vec{u}, i
    \right) \right]} \rho \left( \vec{u} - \vec{v} \right) / \rho \left(
    \vec{u}' - \vec{v} \right) \right] \hspace{1em} \left( \vec{u}' = \vec{u}
    + \widehat{i} \right) \label{ptp}
  \end{equation}
  or
  \begin{equation}
    \tilde{p}_{i, -} = \min \left[ 1, t^{\left[ L_0 - 2 k \left( \vec{u} -
    \hat{i}, i \right) \right]} \rho \left( \vec{u} - \vec{v} \right) / \rho
    \left( \vec{u}' - \vec{v} \right) \right] \hspace{1em} \left( \vec{u}' =
    \vec{u} - \widehat{i} \right) . \label{ptm}
  \end{equation}
  If this happens, all plaquettes in the ladder are flipped and $\vec{u}$ is
  changed to $\vec{u}'$. In the case of rejection the previous configuration
  is kept.
\end{itemize}

\subsection{Complete update sequence}

In this subsection we specialize to $D = 3$ for simplicity. One complete
iteration of algorithm A1 consists of $N_x / L_0$ repetitions of the following
steps
\begin{itemize}
  \item Apply CF to $4 L_0$ cubes around the temporal line at $\vec{u}$. We
  use a helical order and first consider the 4 cubes around the link $\left(
  u, 0 \right)$ with $u_0 = 0$, then at $u_0 = 1$ and so on.
  
  \item This is followed by $n_{\tmop{ps}}$ applications of PS. In all our
  simulations given below we took $n_{\tmop{ps}} = 16$.
\end{itemize}
The cost of one iteration A1 is of the order volume like a conventional sweep.
We here try to mimic the successful worm algorithm in so far as we attach our
update moves to the defect, which now is a (straight) line instead of a point.
A difference in the case at hand is that defect moves alone -- within our
restricted class as discussed before -- are by far not sufficient for
ergodicity. We have mimicked this conglomerate also in the Ising model by
proposing flips of links around only those plaquettes of which a defect forms
a corner. This works but offers no advantage in this case.

In the algorithm just described the defects at $\vec{u}$ and $\vec{v}$ are not
treated on the same footing -- we only shift $\vec{u}$ -- although, due to the
relocation step in PS they can both reach all positions. This is in contrast
to (\ref{Polens}) where they play a symmetric role. Analogous options appear
already for the point defects in spin models, where we found the asymmetric
algorithm easier to program and equally efficient as a manifestly symmetric
variant. We make no such attempt here.

As described before only temporal defect line pairs are included. Their
migrations over the torus allow to change wrapping numbers in the $0 k$
planes, but not in the purely spatial planes (only 12 in $D = 3$). Therefore
A1 simulates an ensemble where one dynamically sums over $\gamma_{12} = \pm 1$
with the other twists trivial.

A second version A2 of the algorithm results, if we allow a pair of Polyakov
lines in any of the $3$ directions, but always only one pair at a time. We
denote by $j^{\left( \mu, \vec{u}, \vec{v} \right)}$ the current with lines in
the $\mu$-direction with $\vec{u}, \vec{v}$ locating these lines. In
(\ref{Polens}) an additional summation over $\mu$ is included,
\begin{equation}
    \mathcal{Z}'  =  \sum_{\mu} \sum_{\vec{u}, \vec{v} \in
    \mathcal{D}_{\mu}} \rho^{- 1}_{\mu} \left( \vec{u} - \vec{v} \right) Z
    [j^{\left( \mu, \vec{u}, \vec{v} \right)}]
\end{equation}
and $\mathcal{D}_{\mu}$ is the $D - 1$ dimensional sublattice of sites $x$
with $x_{\mu} = 0$. In addition the formulae in subsection \ref{PSsubsec} have
to be modified in the obvious way.

The practical difference between A2 and A1 is simply that in PS, when $\vec{u}
= \vec{v}$ is encountered, a new value $\mu' \in \left\{ 0, 1, 2 \right\}$ is
proposed together with a random position of $\vec{u}' = \vec{v}'$ in
$\mathcal{D}_{\mu'}$. The proposal is accepted with the probability $\min
\left( 1, L_{\mu'} / L_{\mu} \right)$. This results in an ensemble as in
(\ref{ZZj}) where all wrapping numbers contribute. If all extensions $L_{\mu}$
are the same the acceptance step is trivial and the resulting histograms of
occurring $\vec{u} - \vec{v}$ values can be combined. If this is not the case,
they are collected separately and yield simultaneous results about Polyakov
loop correlations in several geometries.

\subsection{Rejection free Polyakov shifts}\label{NRsec}

A potential problem with algorithm A1 (and also A2) can be small acceptance
rates for the shifts in PS, if $L_0$ is large. After all we propose nonlocal
albeit only one-dimensional changes to the configuration. Simulating $L^3$
lattices at the critical point, we indeed find acceptances falling with $L$
which reach values of about 1.6\% only for $L = 64$.

In {\cite{Liu:2010uw}} a so-called rejection-free version for general
worm-type algorithms was proposed. If we consider a PS step in A1, then the
total probability that one of the four conceivable moves takes place is given
by
\begin{equation}
  A \left( \vec{u} ; k \right) = \frac{1}{4} \left( \tilde{p}_{1, +} +
  \tilde{p}_{2, +} + \tilde{p}_{1, -} + \tilde{p}_{2, -} \right) .
\end{equation}
Note that $1 - A > 0$ is the probability of keeping $\vec{u}$ unchanged. It is
easy to show {\cite{Liu:2010uw}} that if we replace PS by a step PSnr
\begin{itemize}
  \item if $\vec{u} = \vec{v}$ holds, the same steps as in PS are taken,
  
  \item for $\vec{u} \neq \vec{v}$ one of the four moves discussed before is
  chosen and executed with the normalized probabilities
  \begin{equation}
    p_{i, \pm} = \frac{1}{4 A \left( \vec{u} ; k \right)}  \tilde{p}_{i, \pm}
  \end{equation}
\end{itemize}
then PSnr has the Boltzmann weight in
\begin{equation}
  \mathcal{Z}_{\tmop{nr}} = \sum_{n, \vec{u}, \vec{v}} t^{\sum_{x, \mu < \nu}
  n (x, \mu, \nu)} \rho^{- 1} \left( \vec{u} - \vec{v} \right) \delta
  [\partial^{\ast} n - j^{\left( \vec{u}, \vec{v} \right)}] \left\{
  \delta_{\vec{u}, \vec{v}} + \left( 1 - \delta_{\vec{u}, \vec{v}} \right) A
  \left( \vec{u} ; k \right) \right\} \label{ZZnr}
\end{equation}
as a fixed point. To construct a complete simulation algorithm A1nr we need to
replace beside PS $\rightarrow$ PSnr also the CF step by one, where the
Boltzmann ratio corresponding to (\ref{ZZnr}) is formed for the acceptance
step. Finally also a version A2nr is programmed, where $A \left( \vec{u} ; k
\right)$ is suitably generalized to depend on the direction of the defect
lines. Since in any case the relative weights of {\tmem{vacuum}} graphs are
the same in (\ref{Polens}) and (\ref{ZZnr}) all expectation values
(\ref{mean0}) are as before.

We see that the term `rejection free' refers to the Polyakov shifts in
non-vacuum configurations only. Acceptance rates in CF are however at
efficient levels in all our simulations. If now the remaining acceptance rate
of PSnr out of vacuum gets too small, we may exploit our freedom of choosing
$\rho$ to improve this. For example the choice
\begin{equation}
  \rho \left( \vec{u} - \vec{v} \right) = \delta_{\vec{u}, \vec{v}} + c \left(
  1 - \delta_{\vec{u}, \vec{v}} \right) \label{rhoc}
\end{equation}
with $c < 1$ reduces the relative weight $\rho^{- 1}$ of vacuum configurations
and thus enhances the probability to leave them. A few brief experiments
allowed us to arrange for reasonable acceptance rates for these moves, too. We
find this simple choice for $\rho$ to be a good one at the critical point,
where we will implement A2nr, while in the confined phase in connection with
A1 other choices will be more favorable.

\section{Numerical Simulations}

\subsection{Simulations close to the critical point}

We first report on a number of simulations at
\begin{equation}
  \overline{\beta}_c = 0.76141346 \left( 7 \right) \hspace{1em}
  \Leftrightarrow \hspace{1em} \overline{t}_c = 0.64190876 \left( 4 \right)
  \label{tbarc}
\end{equation}
which corresponds to the dual of the estimate given for the spin model 
in\footnote{In the meantime the precision of the estimate for the
critical temperature has been further improved~\cite{Hasenbusch:2012}.}
{\cite{Deng:2003wv}}.
On each lattice of sizes $L_{\mu} \equiv L = 6, 8, 12,
16, 24, 32, 48$ we have executed $8 \times 10^6$ iterations of A2nr with
$\rho$ from  (\ref{rhoc})  and $c = \left( 8 / L \right)^2$ after a few tests.
All observed acceptance rates of CF are around 50\% and the remaining
acceptance of PS out of vacuum configurations drops from 34\% at $L = 6$ to
23\% at $L = 48$ with these choices.

\begin{figure}[htb]
\begin{center}
  \resizebox{0.7\textwidth}{!}{\includegraphics{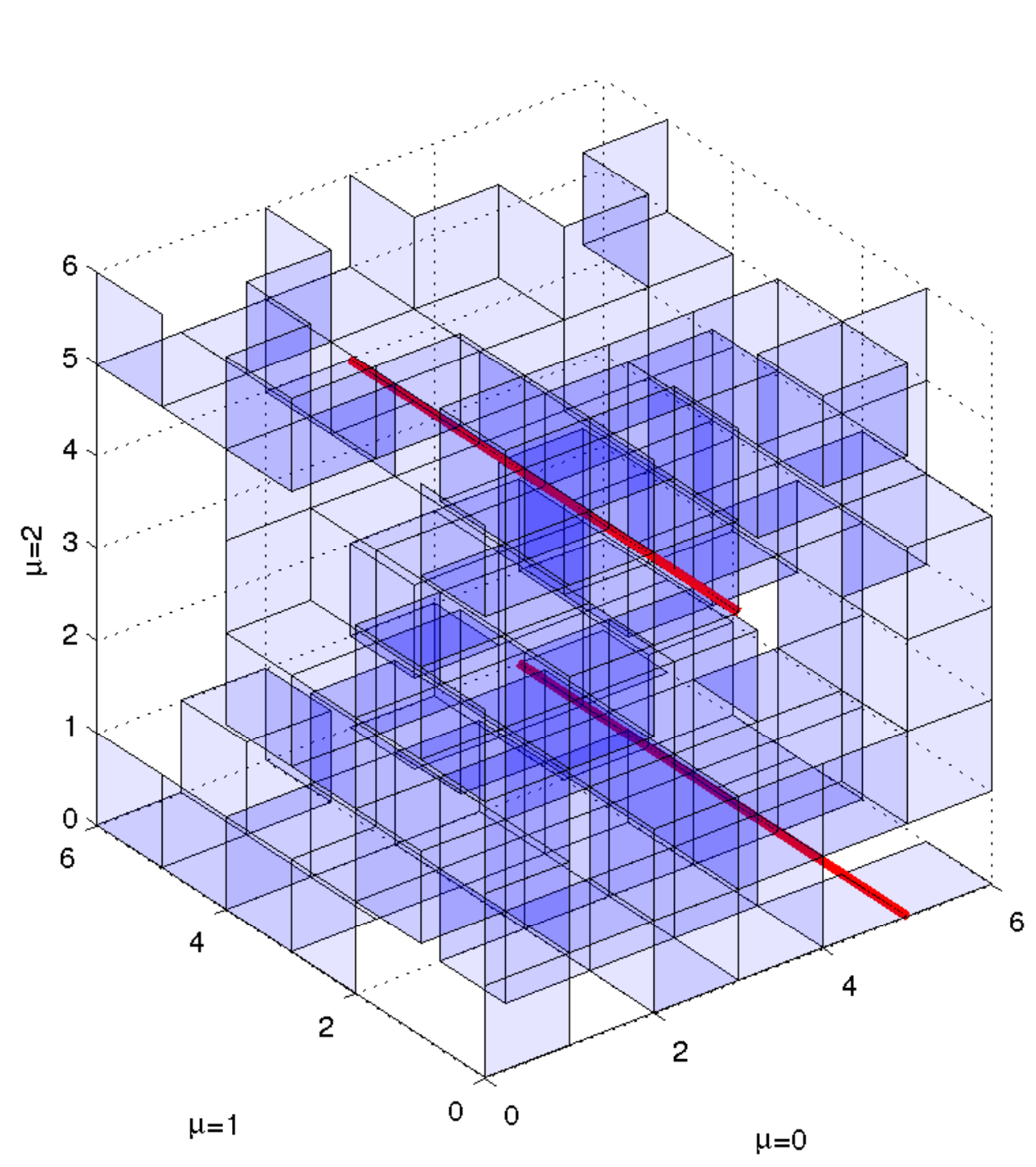}} 
  \caption{A typical configuration on a $6^3$ lattice at $t = \overline{t}_c$.
  The Polyakov line defects are shown as fat red lines.\label{figconf}}
\end{center}
\end{figure}

One of the simplest observables is the average fraction of plaquettes
participating in the random surface of vacuum configurations
\begin{equation}
  \Theta = \frac{1}{N_p} \sum_{x, \mu < \nu}  \langle \langle n \left( x, \mu,
  \nu \right) \rangle \rangle_0
\end{equation}
which is analogous (and equivalent, see (\ref{E0})) to the average plaquette.
We measure values like for example $\Theta = 0.33751 \left( 2 \right)$ at $L =
6$ and $\Theta = 0.33493 \left( 4 \right)$ at $L = 48$. In Fig. \ref{figconf}
we visualize a typical graph occurring in \ the $L = 6$ simulation, larger
systems look even more cluttered. Histograms counting the frequencies of
separations $\vec{u} - \vec{v}$ are accumulated after each individual PS step
and turn out to be rather flat, as expected in the critical theory, where the
Polyakov correlation decays only slowly.

\begin{figure}[htb] \begin{center}
  \resizebox{!}{0.5\textwidth}{\includegraphics{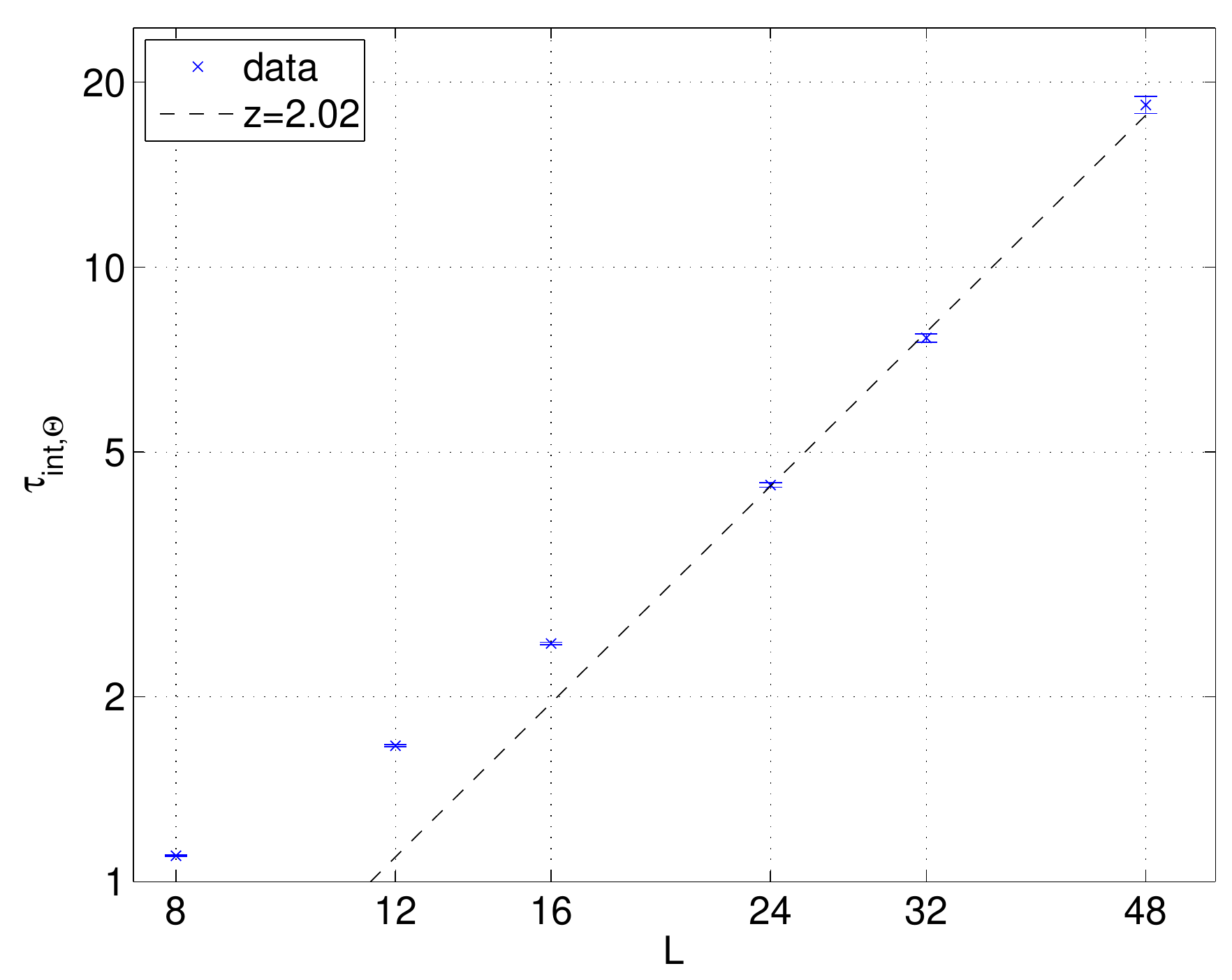}}
  \caption{Double logarithmic plot of integrated autocorrelation times of
  $\Theta$ in units of A2nr iterations which are comparable to
  `sweeps'.\label{figtauint}}
\end{center} \end{figure}

During each iteration of A2nr we separately accumulate contributions to
observables from the subset of vacuum configurations encountered during this
iteration, like for example for $\Theta$. In the end we perform an
autocorrelation analysis of this time series to estimate errors as described
in {\cite{Wolff:2003sm}}. The such defined integrated autocorrelation time of
$\Theta$ is shown in Fig. \ref{figtauint}.

It is in units of A2nr iterations which cost proportional to $L^3$. We see
that while the absolute values are not too large on our lattices the rate of
growth exhibits critical slowing down $\tau_{\tmop{int}} \propto L^z$ with an
effective dynamical exponent for our range of lattice sizes close to two. Note
that we cannot make statements on the truly asymptotic dynamical exponent on
the basis of these data.

\begin{figure}[htb] \begin{center}
  \resizebox{0.9\textwidth}{!}{\includegraphics{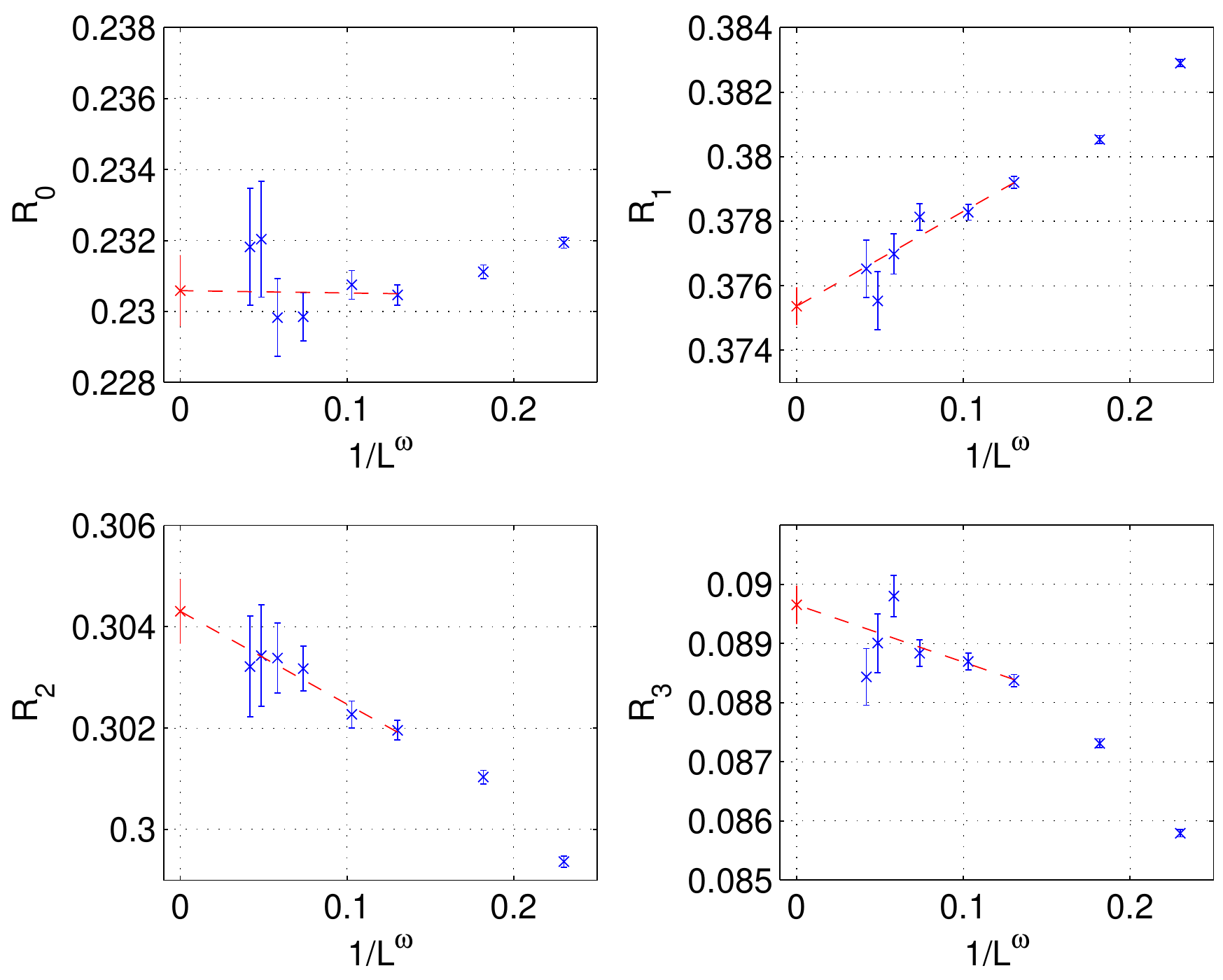}}
  \caption{Plots of $R_n$ versus $L^{- \omega}$ for simulation at $t =
  \overline{t}_c$ of {\cite{Deng:2003wv}}.\label{Rvalues}}
\end{center} \end{figure}

Of greater interest than $\Theta$ are the topological observables
(\ref{wdef}). Taking into account the symmetries between all planes we here
measure the set of observables
\begin{equation}
  R_n = \langle \left\langle \delta_{n, w_{01} + w_{02} + w_{11}}
  \right\rangle \rangle_0, \hspace{1em} n = 0, 1, 2, 3. \label{Robs}
\end{equation}
Due to (\ref{wgamma}) each $R_n$ can be simply related to ratios of partition
functions with twisted and untwisted boundary conditions and is hence expected
to possess finite continuum or scaling limits. In particular, at the exact
critical point there are definite finite values $R_n^{\ast}$ for each $R_n$ in
the limit $L \rightarrow \infty$ at $t = t_c$. In fact, this property may be
used to determine $t_c$. Note that the values $R_n$ are not independent, of
course, but have to sum to unity. Our raw results on these quantities are
compiled in Table \ref{Rtab} in the Appendix. We here consider the expected
finite size scaling behavior close to the critical point in the form
\begin{equation}
  R_n \left( t_c, L \right) = R_n^{\ast} + \alpha_n L^{- \omega} + \ldots,
  \label{FSSR}
\end{equation}
where $\omega$ is the exponent of the leading scaling violations. In our
simulations at $t = \overline{t}_c$ we find the results shown in Fig.
\ref{Rvalues}. In the variable $L^{- \omega}$ a linear approach to
$R_n^{\ast}$ is thus expected. Recent determinations of the exponent $\omega$
can be found in {\cite{Deng:2003wv}} and {\cite{Hasenbusch:2010}} - we use
$\omega = 0.82$.

We show simple fits for our lattices with $L \geqslant 12$. Details on raw
data are given in the appendix. We here just list the extrapolated values
$R_0^{\ast} = 0.2306 (10)$, $R_1^{\ast} = 0.3754 (6)$, $R_2^{\ast} = 0.3043
(6)$, $R_3^{\ast} = 0.0897 (3)$.

\

\subsection{Simulations in the confined phase}

In the confined phase $\beta < \beta_c$ we simulate with algorithm A1, such
that the distribution of $\vec{u} - \vec{v}$ has a direct relation to the
temporal Polyakov loop correlator (\ref{Gpol}). We do not generate surfaces
wrapping around the 12 plane corresponding to dynamically summing over twisted
and untwisted boundary conditions for this plane. The Polyakov correlator is
however periodic in space in this situation.

For our test we took the coupling $\beta = 0.73107$ with $L_1 = L_2 = L = 64$
and a series of inverse temperatures $L_0 = 6, 8, \ldots, 26, 28.$ Such values
(but smaller $L_0$) have been adopted in {\cite{Caselle:2002ah}} and a string
tension in lattice units of $\hat{\sigma} = 0.0440 \left( 3 \right)$ has been
estimated based on earlier literature. In our simulations in the confined
phase we want to use a $\rho \left( \vec{x} \right)$ in (\ref{Polens}) that
essentially captures the decay of the two point function (\ref{Gpol}).
Inspired by the lowest term in (\ref{GLW}) we prepare\footnote{
The sum is singular for $x_1=x_2=0$. We replace it by an extrapolation
along the diagonal solving $f(0,0)/f(1,1)=f(1,1)/f(2,2)$.
}
\begin{equation}
  \rho \left( \vec{x} \right) \propto \sum_{\vec{k} \in \mathbb{Z}^2} K_0
  \left[ \hat{M} \sqrt{\left( x_1 - k_1 L \right)^2 + \left( x_2 - k_2 L
  \right)^2} \right], \hspace{1em} \rho ( \vec{0} ) = 1.
\end{equation}
The mass $\hat{M}$ is estimated from $\hat{\sigma}$ by using the first three
terms on the right hand side of $\left( \ref{sigex} \right)$. The sum over
integer $\vec{k}$, which makes $\rho$ periodic, is actually truncated after a
few terms as the remainder would be insignificant compared to roundoff due to
the exponential decay of the Bessel function. We emphasize that any
imperfection in $\rho$ is related to the efficiency of the simulations but
does not entail any systematic error.

\begin{figure}[htb] \begin{center}
  \resizebox{0.7\textwidth}{!}{\includegraphics{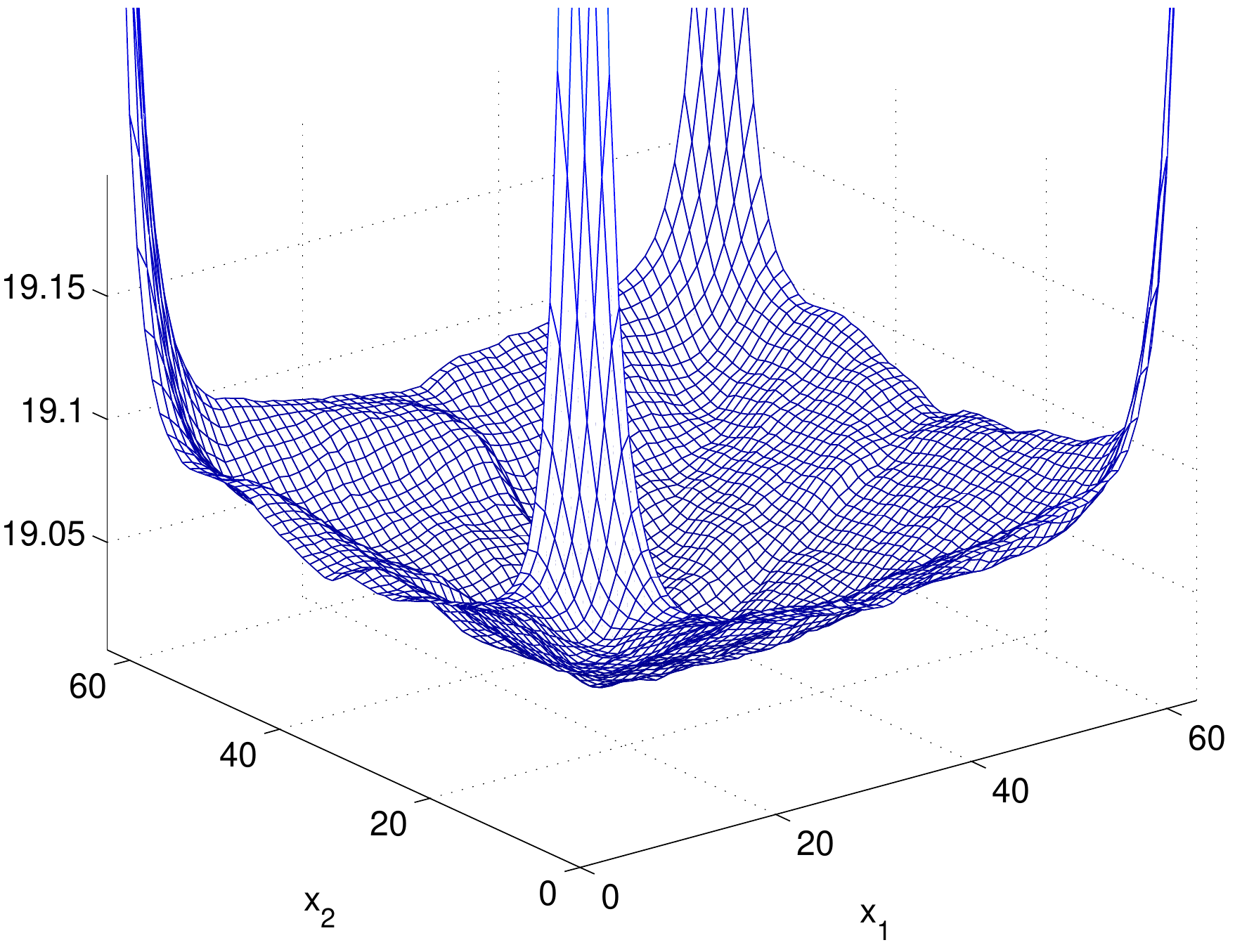}}
  \caption{Log of the number of occurrences of separations $\vec{u} - \vec{v}$
  of the Polyakov defect lines for the run $24 \times 64^2$.\label{Hhist}}
\end{center} \end{figure}

\begin{figure}[htb] \begin{center}
  \resizebox{0.8\textwidth}{!}{\includegraphics{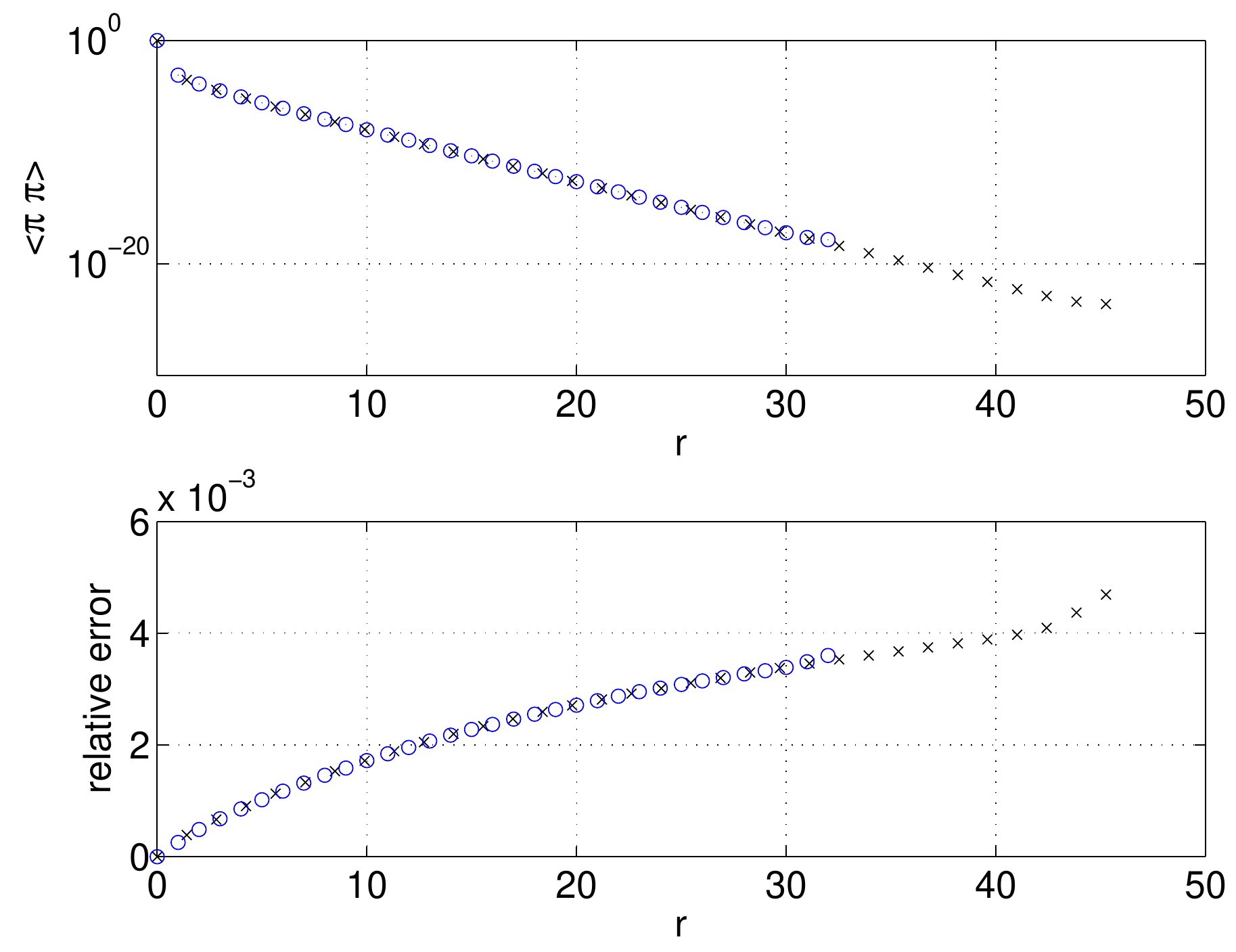}}
  \caption{Polyakov line correlation function (\ref{Gpol}) for the $24 \times
  64^2$ lattice versus Euclidean separation $r = \sqrt{x_1^2 + x_2^2}$.
  Circles refer to on-axis separations, crosses to diagonal ones. Since errors
  are invisible in the upper plot, we separately show the slowly growing
  {\tmem{relative}} errors in the lower plot.\label{Gpolfig}}
\end{center} \end{figure}

For each value of $L_0$ we have performed $1.2 \times 10^7$ iterations of A1.
In Fig. \ref{Hhist} we show the histogram of the frequencies of sampled
separations of the two defect lines in our run for $L_0 = 24$. With the
exception of very short torus distances we find a rather flat behavior (note
the fine scale of this logarithmic plot) which leads to very uniform bin
heights, which, up to correlation effects, would naively imply constant
relative errors. Via (\ref{Gpol}) this allows to measure the correlator which
for the same run is plotted in Fig. \ref{Gpolfig}. To determine errors here,
we had to store the time series for these separations rather than just the
histogram. We see that the exponential decay can be followed over the whole
lattice with relative errors staying small. The plot jointly exhibits `on
axis' and `diagonal' separations which form a very smooth curve and thus
exhibit small violations of rotation invariance. The spatial correlation
length is of order one and therefore the periodicity is just barely visible
close to $r = L / 2$ and $r = L / \sqrt{2}$.

\begin{table}[htb] \begin{center}
  \begin{tabular}{|l|l|l|l|l|l|}
    \hline
    $L_0$ & $\tilde{E}_0$ & $\left| v_0 \left| \right. \right.$ & $L_0$ &
    $\tilde{E}_0$ & $\left| v_0 \left| \right. \right.$\\
    \hline
    6 & 0.160286(15) & 0.165454(28) & 18 & 0.762774(77) & 0.104107(89)\\
    \hline
    8 & 0.279241(21) & 0.138969(31) & 20 & 0.853968(98) & 0.10092(12)\\
    \hline
    10 & 0.384096(28) & 0.126905(37) & 22 & 0.94450(12) & 0.09807(13)\\
    \hline
    12 & 0.482528(36) & 0.118760(45) & 24 & 1.03480(14) & 0.09571(15)\\
    \hline
    14 & 0.577660(47) & 0.112776(55) & 26 & 1.12416(16) & 0.09329(18)\\
    \hline
    16 & 0.670782(60) & 0.108026(60) & 28 & 1.21429(19) & 0.09154(21)\\
    \hline
  \end{tabular}
  \caption{Closed string mass gaps $\tilde{E}_0$ and matrix element $v_0$, see
  (\ref{Ccor}), on $L_0 \times 64^2$ lattices at $\beta = 0.73107$. Each data
  point derives from $1.2 \times 10^7$ iterations of A1.\label{Mtortab}}
\end{center} \end{table}

\begin{figure}[htb] \begin{center}
  \resizebox{0.7\textwidth}{!}{\includegraphics{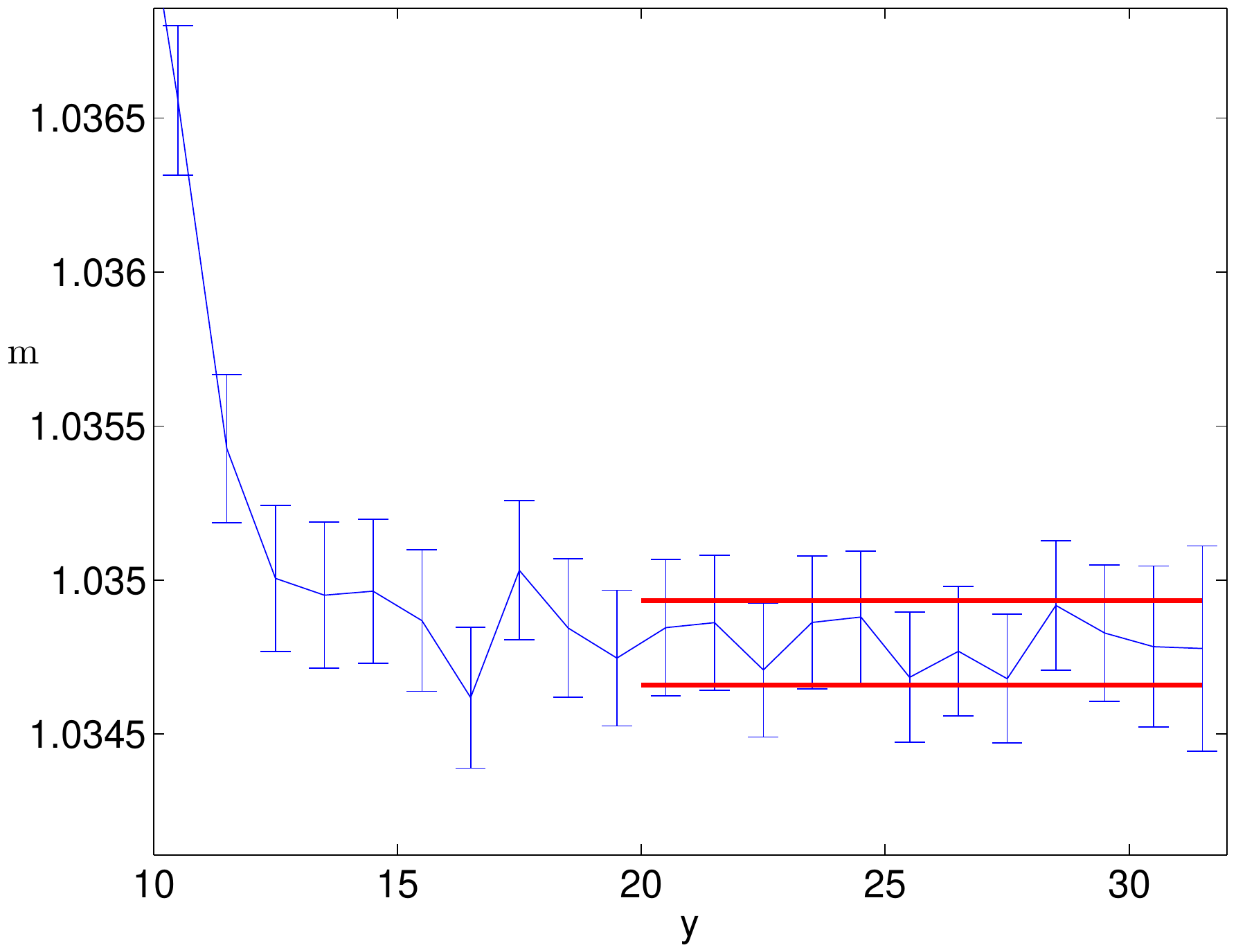}}
  \caption{Effective mass in the run with $L_0 = 24$. \label{meff}}
\end{center} \end{figure}

The closed string mass gap $\tilde{E}_0$ is estimated from (\ref{Cz}) and our
rather precise results are compiled in Table \ref{Mtortab}. A typical case of
the determination of the mass gap is shown in Fig. \ref{meff}. Effective
masses $m \left( y + 1 / 2 \right)$ are determined by solving $C \left( y
\right) / C \left( y + 1 \right) =$ $\cosh \left( m \left( y - L / 2 \right)
\right) / \cosh \left( m \left( y + 1 - L / 2 \right) \right)$. The horizontal
lines show the error band from a fit deeply in the plateau region which leads
to the entries for $\tilde{E}_0 $ in the table. Observed autocorrelation times
$\tau_{\tmop{int}}$ for these derived quantities {\cite{Wolff:2003sm}} range
from 0.508(4) ($\left. L_0 = 6 \right)$ to 0.598(6) ($\left. L_0 = 28
\right)$. The remarkable feature here is, that the errors in the effective
masses {\tmem{do not grow with separation}}. This is due to our judicious
adaptation of the simulated ensemble by choosing $\rho$. The situation is
analogous to the one in the Ising spin model {\cite{Wolff:2008km}} where the
reasons for this success are discussed in more detail.

What is the implication of our fluctuating boundary conditions in the 12 plane
for the correlation (\ref{Cz})? The falloff in the direction $x_1$ is
associated with the transfer matrix in this direction which acts on
wave-`functionals' $\psi \left[ \sigma_0 \left( x_0, x_2 \right), \sigma_2
\left( x_0, x_2 \right) \right]$. The Polyakov operator (\ref{Polop}) acts on
them by multiplication. The integration over links $\sigma_1$ in the Euclidean
theory implies the inclusion of a projector on gauge invariant states in the
thermal trace given by the path `integral'. This refers to gauge
transformations that are periodic in both directions. One may however in
addition consider transformations {\cite{'tHooft:1979uj}} that are
topologically nontrivial, here antiperiodic in $x_2$, and physical states may
be even or odd under them. By summing over twisted and untwisted boundary
conditions in the 12 plane we include a projector also with respect to this
quantum number. Under the assumption that the ground state is even, such a
projection is no disadvantage.

\begin{figure}[htb] \begin{center}
  \resizebox{0.7\textwidth}{!}{\includegraphics{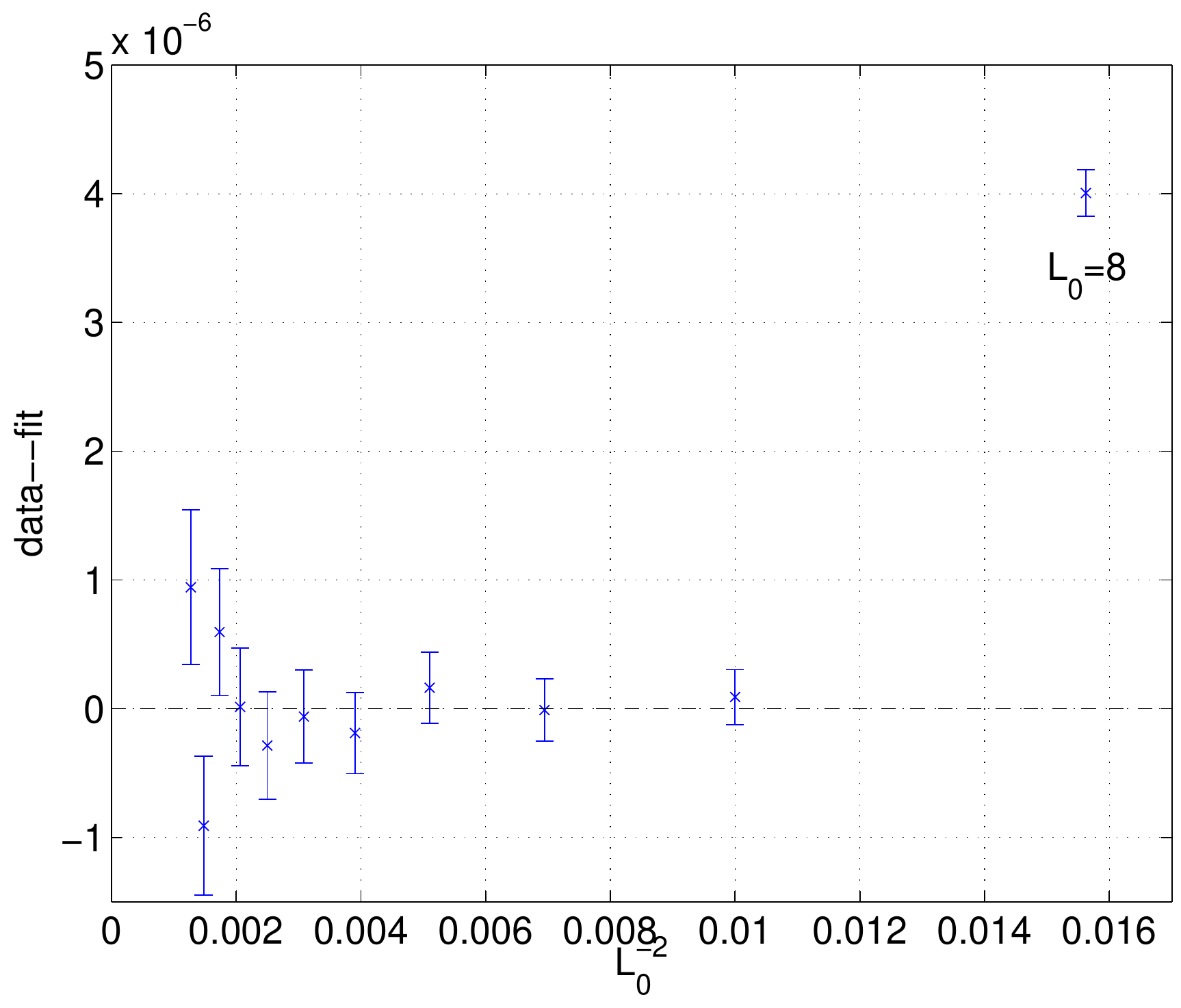}}
  \caption{Difference between data of Table \ref{Mtortab} and the second fit
  of the form (\ref{NGfit}) listed in Table \ref{NGfits}.\label{da-fitfig}}
\end{center} \end{figure}

To analyze our data we perform a fit of the form
\begin{equation}
  \frac{\tilde{E}_0^2}{L_0^2} = c_0 - c_1 \frac{1}{L_0^2} . \label{NGfit}
\end{equation}
To have visible errors at all we immediately plot the difference between the
left and right hand side of such a fit against $L_0^{- 2}$ in Fig.
\ref{da-fitfig}. The fit here was derived from the subset $L_0 \geqslant 12$
of the data. The point with $L_0 = 8$ is much higher while $L_0 = 6$ is far
off the panel. At these time extents the low temperature expansion has clearly
and rather abruptly \ broken down. In fact, for our $\beta$ the phase
transition is close to $L_0 \approx 4$ according to {\cite{Caselle:2002ah}}.

We note that (\ref{NGfit}) is exactly equivalent to the Nambu Goto form
(\ref{NGrel}) if we identify
\begin{equation}
  \tmop{Nambu} \tmop{Goto} : \hspace{2em} c_0 = \sigma^2, \hspace{1em} c_1 =
  \frac{\pi}{3} \sigma \hspace{1em} \Rightarrow \hspace{1em} r =
  \frac{9}{\pi^2} \frac{c_1^2}{c_0} = 1.
\end{equation}
In Table \ref{NGfits} we list a number of fits (\ref{NGfit}) with
{\tmem{free}} $c_0, c_1$ together with the resulting ratio $r$ for whose error
the correlation between $c_0$ and $c_1$ is taken into account.

\begin{table}[htb] \begin{center}
  \begin{tabular}{|l|l|l|l|l|}
    \hline
    $L_{0, \min}$ & $\chi^2 / \tmop{dgf}$ & $r$ & $\sqrt{c_0}$ & $c_1$\\
    \hline
    10 & 8.0/8 & 1.0109(13) & 0.0440330(25) & 0.046363(35)\\
    \hline
    12 & 8.0/7 & 1.0114(23) & 0.0440334(32) & 0.046374(62)\\
    \hline
    14 & 8.0/6 & 1.0111(40) & 0.0440332(42) & 0.04637(11)\\
    \hline
  \end{tabular}
  \caption{Fits (\ref{NGfit}) where $L_0 < L_{0, \min}$ are
  omitted.\label{NGfits}}
\end{center} \end{table}

We see a small but significant deviation of $r$ from one by only about 1\%.
The most likely explanation for this are cut off effects in our opinion, given
that the string tension $\sigma$ in lattice units, which may be taken as
indicative for their size, is about $0.044$. It remains to be discussed to which
terms in the expansion (\ref{E0ex}) our fits are actually sensitive. The
errors of our $\tilde{E}_0$ are 7\%, 24\% , 68\% of the last term in
(\ref{E0ex}) for $L_0 = 10, 12, 14$. The analogous numbers for the next term
are 50\%, 236\%, 897\%. This next term reads $5 \pi^4 / \left( 10368 \sigma^3
L_0^7 \right)$ and this is only taken as a model of conceivable next order
terms. We thus have to conclude that in spite of our rather high precision we
are just still sensitive to the last (known) universal term of order $L_0^{-
5}$ and cannot confirm or disprove the coefficient of $L_0^{- 7}$ here.

In future simulations it will be desirable to simulate smaller lattice
spacings to check the size of cutoff effects and possibly to enlarge the
statistics to the point of seriously probing another term in (\ref{sigex}).

\section{Conclusions}

We have attempted to generalize the worm algorithm for the Ising model
to gauge theories with Ising link variables. The labeling of all-order
(in $\tanh \beta$) strong coupling graphs as configurations of constrained
plaquette fields was straight forward. The essential method of \cite{prokofev2001wacci}
to achieve efficient updates of the graphs was to allow for a pair of point
defects. The possible defects in the gauge case form a very large set of generalized loop
networks. We could not identify among them a suitable subset that strongly reduces
critical slowing down while staying `close to' the vacuum or simple Wilson loop
defects.

The defects in the spin model allowed at the same time for a very precise
estimation of the fundamental correlation at large distance \cite{Wolff:2008km}. We successfully
have generalized this aspect to the gauge model and could compute the Polyakov
line correlator with similar precision. At the same time, the algorithmic efficiency
is rather high on large lattices in three dimensions in spite of critical slowing down.

Other Abelian gauge theories have been considered in the surface representation,
see for example  \cite{Endres:2006xu}, \cite{Korzec:2010sh}, \cite{Gattringer:2012jt}.
The techniques for correlators in this paper can clearly be generalized to these
cases, i.e. to the groups Z($N$) or U(1).
Theories with SU($N$) variables, both gauge and the principal chiral spin 
models\footnote{
Note that while there is mention of worm simulations of SU(3) spin models in the
literature \cite{Mercado:2012ue}, their nontrivial symmetry is Z(3).}
are open problems in this context to which we hope to come back in the future.

{\noindent}\tmtextbf{Acknowledgments}: UW is indebted to many people for
helpful discussions: Tim Garoni, Martin Hasenbusch, Stefano Lottini, Mike
Peardon, Stefan Sint, Erhard Seiler, Rainer Sommer and Peter Weisz. Financial
support of the DFG via SFB transregio 9 is acknowledged.

\section*{Appendix} \label{appa}

In this appendix we list the raw data on which Fig. \ref{Rvalues} is based.
Each line represents $8 \times 10^6$ iterations of algorithm A2nr as described
in subsection \ref{NRsec} except for $L = 48$ where the number has been
doubled. The error of these quantities seems to grow roughly proportionally to
$L^{1.2}$ with a fixed number of iterations whose costs scale like $L^3$.

\begin{table}[htb] \begin{center}
  \begin{tabular}{|l|l|l|l|l|}
    \hline
    $L$ & $R_0$ & $R_1$ & $R_2$ & $R_3$\\
    \hline
    6 & 0.23194(15) & 0.38290(10) & 0.29937(11) & 0.08579(6)\\
    \hline
    8 & 0.23112(19) & 0.38053(13) & 0.30103(13) & 0.08731(7)\\
    \hline
    12 & 0.23047(29) & 0.37920(18) & 0.30196(19) & 0.08838(10)\\
    \hline
    16 & 0.23075(40) & 0.37828(25) & 0.30227(26) & 0.08870(14)\\
    \hline
    24 & 0.22985(68) & 0.37813(41) & 0.30317(44) & 0.08884(23)\\
    \hline
    32 & 0.22983(110) & 0.37699(62) & 0.30338(69) & 0.08980(35)\\
    \hline
    40 & 0.23204(163) & 0.37553(90) & 0.30343(100) & 0.08901(50)\\
    \hline
    48 & 0.23182(165) & 0.37653(89) & 0.30321(100) & 0.08844(48)\\
    \hline
  \end{tabular}
  \caption{Results for the observables (\ref{Robs}) from simulations on $L^3$
  lattices with periodic boundary conditions in all directions at the
  estimated critical $\beta$ given in (\ref{tbarc}).\label{Rtab}}
\end{center} \end{table}

\end{document}